\documentclass[a4paper,11pt]{article}
\usepackage{jinstpub} 
\usepackage{subcaption}
\usepackage{booktabs}
\usepackage{svg}

\title{PocketWATCH: Design and operation of a multi-use test bed for water Cherenkov detector components in pure and gadolinium loaded water}

\author[a,1]{M. Thiesse\note{Corresponding author.},}
\author[a]{S. T. Wilson,}
\author[a]{J. Fannon,}
\author[a,2]{M. Malek,\note{Now at BanterBots.ai.}}
\author[a,3]{J. McElwee,\note{Now at LP2i Bordeaux.}}
\author[a]{A. Scarff,}
\author[a]{and L. F. Thompson}

\affiliation[a]{Department of Physics and Astronomy,\\
University of Sheffield, Hounsfield Road, Sheffield S3 7RH, U.K.}

\emailAdd{m.thiesse@sheffield.ac.uk}

\abstract{The PocketWATCH facility is a unique multi-purpose test bed designed to replicate the conditions of large water Cherenkov detectors. Housed at the University of Sheffield, the facility consists of a light-tight 2000~L ultrapure water tank with purification and temperature control systems. Water temperature, resistivity, and UV attenuation in the tank are monitored and shown to be stable over time. The system is also shown to be compatible with a solution of 0.2\% gadolinium sulfate, allowing further utility in testing equipment bound for the next generation neutrino and nucleon decay water Cherenkov particle detectors. The relevant water quality parameters are shown to be stable whilst running in Gd-mode, thereby providing a suitable test bed for hardware development in a realistic, ex situ environment.}


\begin{document}
\maketitle
\flushbottom

\section{Introduction}
\label{sec:intro}

Large pure water Cherenkov particle detectors have long been a staple of neutrino and nucleon decay experiments. Experiments such as, IMB~\cite{BECKERSZENDY1993363}, Kamiokande~\cite{kajita2012origin}, Super-Kamiokande (SK)~\cite{FUKUDA2003418} and the Sudbury Neutrino Observatory~\cite{BOGER2000172} have historically led these areas of physics in terms of sensitivity and discovery, with a large part of their success attributable to their ability to operate stably in an ultra-pure water environment. The next-generation flagship ultra-pure water Cherenkov neutrino and nucleon decay experiment, Hyper-Kamiokande (HK)~\cite{protocollaboration2018hyperkamiokande}, is currently under construction in Kamioka, Japan. The design of HK is influenced by a great deal of research and development into new technologies to improve the detector performance and reduce systematic uncertainties, beyond what has ever been achieved. A proposed operating mode of HK is to follow in the footsteps of Super-Kamiokande Gd~\cite{ABE2022166248}, where the ultra-pure water is loaded with gadolinium to significantly improve the neutron tagging capability and allow discrimination of neutrinos and antineutrinos.

Important factors under consideration in detector R\&D include the dynamic calibration of optical properties such as scattering and absorption, detector construction materials compatibility with ultra-pure and gadolinium-loaded water, and investigation of photomultiplier noise in water as a function of temperature and water quality. Such work requires access to a test bed facility in which to perform measurements and calibrations of proposed hardware designs. The facility should be realistically similar to the operating conditions of HK in that the water is ultra-pure (>18~M$\Omega\cdot$cm) and temperature-controlled, and the environment is dark enough to operate photomultipliers in single-photoelectron mode. In this paper, we describe the design, operation, and performance of PocketWATCH (Pocket WATer CHerenkov), a multi-purpose test bed at the University of Sheffield for the development of detector components for HK and other proposed and future water Cherenkov facilities, such as the Boulby Underground Technology Testbed for Observing Neutrinos (BUTTON)~\cite{knealeButton}. 

\section{Facility Description}
\label{sec:facility}

\begin{figure}[htb]
\centering
\begin{minipage}[t]{0.56\textwidth}
    \centering
    \includegraphics[width=\textwidth]{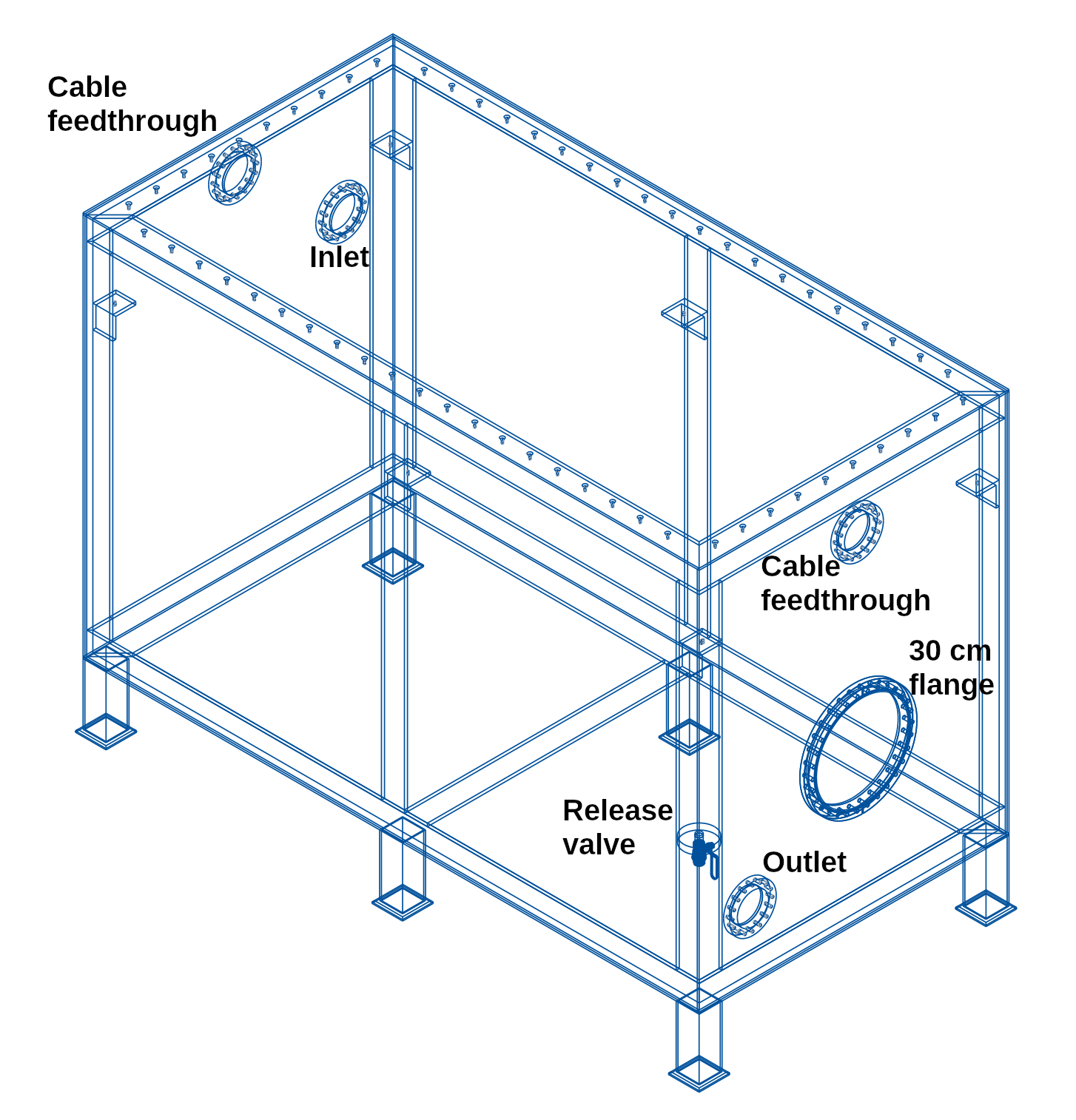}
    \caption{\label{fig:tankdrawing}CAD rendering of the PocketWATCH tank.}
    \end{minipage}
    \hfill
    \begin{minipage}[t]{0.42\textwidth}
    \centering
    \includegraphics[width=\textwidth]{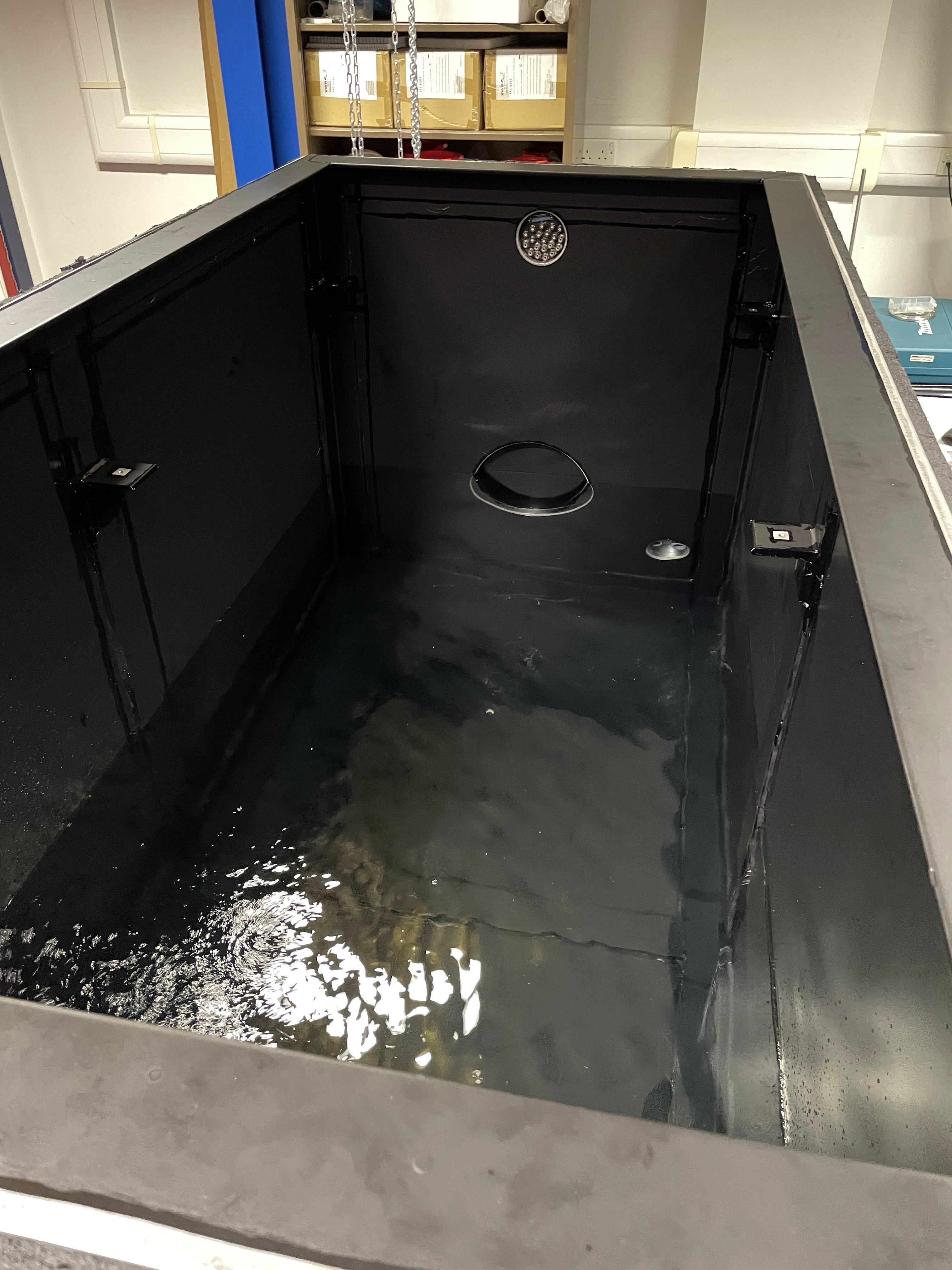}
    \caption{\label{fig:duringfilling}PocketWATCH photo, taken without lid and approximately halfway through filling with pure water. The lid is removed and set aside for the sake of this photo.}
    \end{minipage}
\end{figure}

PocketWATCH is a $2\times2\times1.25$~m tank (Figs.~\ref{fig:tankdrawing} and \ref{fig:duringfilling}) containing 2000~L of water, with 25~cm of air gap above the water level for \textit{in situ} control electronics and gantry systems. The tank is raised 20~cm off the floor on six legs and is constructed of 6~mm thick SS316 plates. A release valve is installed on one of the bottom square plates to facilitate draining when required. A 3~mm thick SS316 lid sits on top of the tank walls and compresses a 1~cm thick foam gasket, forming a hermetic seal and allowing rejection of ambient light. A 2-tonne overhead crane and a hand-pulled manual crane allow objects to be stably lowered into and raised out of the tank.

The tank includes four 10 cm flanges; two above the water level for sensors, signals, power and fibre optics feedthroughs and two below the water level at opposite diagonal corners, containing separate ports for water supply, filtered and chilled, and return to the loop. A 30~cm port was included on one tank wall to facilitate a proposed future experiment to take place in the facility.

A three-axis gantry is available for installation at the water surface inside the tank. The x- and y-axes move a platform to almost any point on the water surface, which carries a platform that can be lowered via pulleys to any height within the tank. The purpose of such a system is to allow the automatic positioning of light sources or detectors around test equipment mounted underwater.

To ensure the inside of the tank is sufficiently dark to operate photomultipliers in single-photon counting mode, several light-tightness measures are implemented. The tank walls are covered in 3M 2080 Matt Deep Black vinyl to suppress light reflections. Also, black water-filled polypropylene balls rest on the surface of the water and provide 91\% surface coverage through optimal hexagonal circle packing, suppressing light reflections from the water surface. Thick-walled ABS pipes which carry water to and from the tank, and the thick foam gasket between the tank walls and the lid, are shown to sufficiently block ambient light from entering the tank. Finally, vacuum-compatible flanges and electronic feedthroughs ensure that no stray light is allowed into the tank volume.

\begin{figure}[htb]
    \centering
    \includegraphics[width=\textwidth]{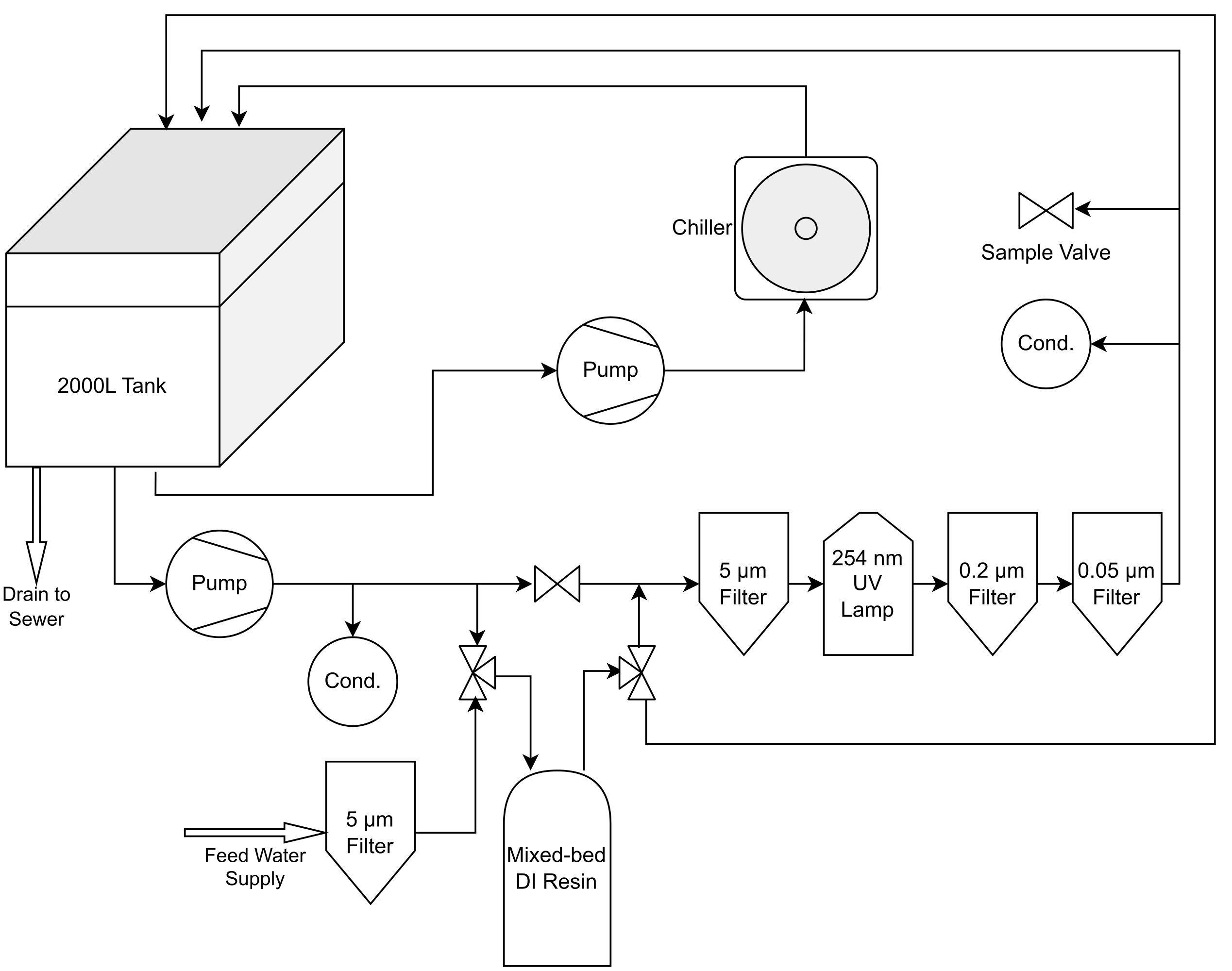}
    \caption{\label{fig:filtration}Schematic of the PocketWATCH filtration system.}
\end{figure}

An ultra-pure water filtration system was designed and supplied by Veolia Water Technologies to provide 18.18~M$\Omega\cdot$cm resistivity at 25 $^\circ$C water at 20~L~min$^{-1}$ flow rate. At this flow rate, the idealised turnover time is approximately 1.7~hours. The filtration system and the tank water are in constant recirculation during operation, so the water is continuously polished. A schematic of the system is shown in Fig.~\ref{fig:filtration}. Dissolved anions and cations, such as Ca$^{2+}$, Cl$^{-}$ and the products of CO$_2$ dissolution including bicarbonate, carbonate and carbonic acid, are removed via the nuclear-grade mixed-bed deionisation (DI) resin cylinder. Suspended particulates are filtered through successively smaller filters, down to 0.05~$\mu$m pore size. Bacterial growth is suppressed by the 254~nm UV sterilisation lamp. Pre- and post-filtration resistivity and water temperature are measured using dedicated sensors (Mettler Toledo UniCond 2-electrode with cell constant 0.1~cm$^{-1}$).

There are four main operating modes of the filtration system:
\begin{enumerate}
    \item ``Normal Operation, Pure Water'' -- Water is drawn from the tank via the pump, sent through the DI cylinder, and then through the UV lamp and particulate filters before returning to the tank.
    \item ``Normal Operation, Gd Water'' -- Water is drawn from the tank via the pump, bypassing the DI cylinder, then sent through the UV lamp and particulate filters before returning to the tank. 
    \item ``Fill and Recirculate'' -- The feed water supply passes through the preliminary 5~$\mu$m filter before passing through the DI cylinder and into the tank via the alternate return line. Separately, the tank water is pumped through the remainder of the filtration loop -- bypassing the DI cylinder, passing through the UV lamp and particulate filters before returning to the tank via the main return line.
    \item ``Sanitisation'' -- Recirculation of sanitisation chemical through all lines of filtration system, with filter and DI media removed.
\end{enumerate}

In a separate recirculation loop from the water filtration, an Applied Thermal Control Ltd. K4 4.5~kW chiller is used to maintain the water temperature in the range $4-30~^\circ$C to within $\pm~0.1~^\circ$C at 17~L~min$^{-1}$. In addition to the standard vapour compression and expansion cooling process, the chiller includes a hot gas bypass valve to allow heating of the water if the temperature setpoint is higher than the inlet temperature.

Once per year, the facility undergoes necessary maintenance. All four particulate filters and the UV lamp are removed and the system is fully sanitised. A chemical sanitiser is used which contains hydrogen peroxide, acetic acid and peracetic acid (e.g. Hydrex 7110). The sanitiser is maintained in the tank and recirculation loops at a concentration of 0.8\% w/w for a minimum of one hour contact time with all parts of the filtration system and chiller loops. Once completed, the sanitiser is flushed, filters and UV lamp replaced, and PocketWATCH is returned to service.

\section{System Performance}
\label{sec:purewater}

To perform as a test bed for calibration hardware and testing, PocketWATCH must exhibit high levels of stability, purity and light-tightness. The system is specifically designed to achieve these goals, with several metrics.

\subsection{Temperature Stability}

Whilst the K4 chiller is specified to maintain an output water temperature to within 0.1 $^\circ$C of the setpoint, there is a very large total surface area through which heat can flow into and out of the main water volume. To ensure the bulk water temperature can still be maintained, the tank was lagged with 19~mm thick Armaflex Class O insulation with thermal conductivity less than 0.033~W/(m$\cdot$K) at 0 $^\circ$C. All long lengths of pipework between the chiller and tank, and filtration system and tank, are insulated with foam as well. Finally, the plant room is itself air conditioned, held at 20 $^\circ$C year round.

\begin{figure}[htb]
    \centering
    \includegraphics[width=0.7\textwidth]{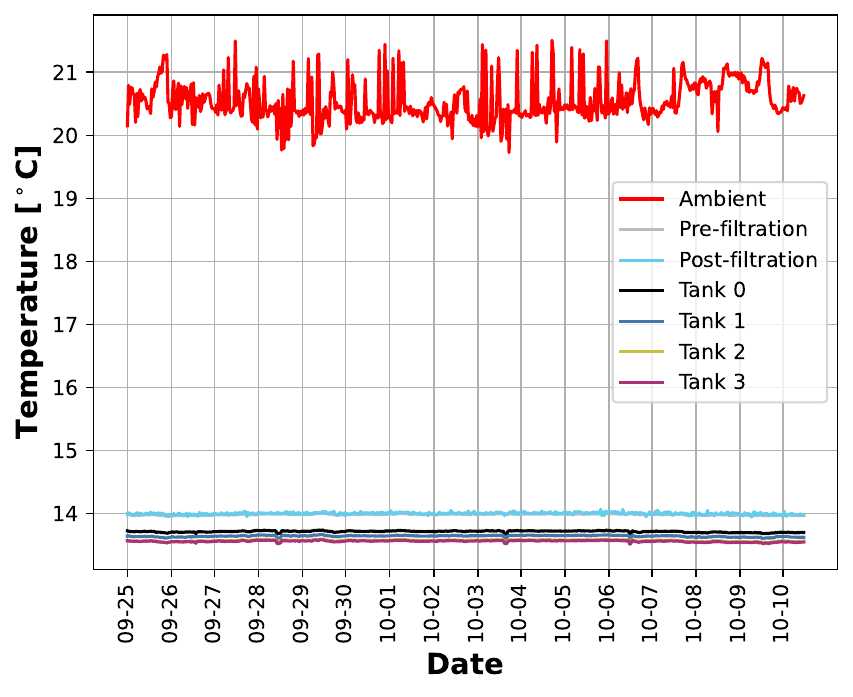}
    \caption{\label{fig:temperature}Temperatures of different parts of the PocketWATCH facility over the course of 16 days of pure-water running. The ambient temperature is the laboratory air temperature, and other measurements are at different points in the water system.}
\end{figure}

The result of these measures is that the bulk water is held stably to within $0.012~^\circ$C for a setpoint of $14^\circ$C. A small offset in measured temperature between different probes is most likely caused by slight differences in ADC calibration of the RTD signal conditioner and transmitter. A plot of the temperature stability, as measured by a series of Pt100 RTD sensors distributed throughout the tank volume, filtration system, and laboratory is shown in Fig.~\ref{fig:temperature}.

\subsection{Resistivity}

The resistivity of the ultra-pure water is constantly monitored pre- and post-filtration. Since there is no \textit{in situ} resistivity instrumentation in the bulk tank volume, the two measurement points are taken as proxies for the minimum and maximum purity of the tank water. Due to the dissolution of atmospheric gases and leeching of chemicals from various parts of the tank, filtration system, chiller, and any other components submerged inside the tank, the resistivity of the water post-filtration is always the maximum purity of the bulk water. The pre-filtration resistivity is assumed to be the minimum water purity at all points within the tank so long as the assumption of efficient impurity mixing throughout the entire tank is valid.

\begin{figure}[htb]
    \centering
    \includegraphics[width=0.7\textwidth]{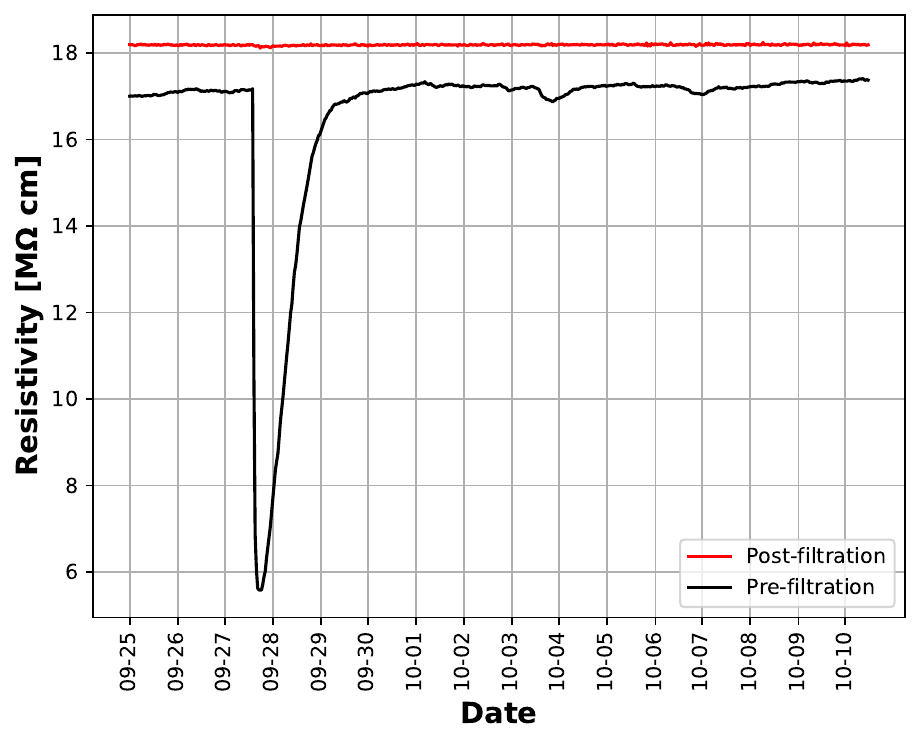}
    \caption{\label{fig:resistivity}Resistivity (compensated to $25~^\circ$C) of pure tank water. The large dip in pre-filtration resistivity occurred following a brief, 20~minute period when the tank lid was removed then replaced.}
\end{figure}

Fig.~\ref{fig:resistivity} shows a trace of the pre- and post-filtration resistivity (compensated to $25~^\circ$C) while running in pure-water mode. The filtration system constantly maintains a supply of ultra-pure water (18.18~M$\Omega\cdot$cm), and observes a small drop in resistivity on return from the tank volume (17~M$\Omega\cdot$cm). To illustrate the effect of atmospheric gas dissolution on the water resistivity, during this particular measurement period a large drop in pre-filtration resistivity occurred on 27$^{th}$ September when the tank lid was removed for approximately 20 minutes. Once the lid was replaced, the resistivity once again returned to the prior value after several hours of recirculation. 

\subsection{UV Light Transmission}

\begin{figure}[htb]
\centering
\begin{minipage}[t]{0.49\textwidth}
    \centering
    \includegraphics[width=\textwidth]{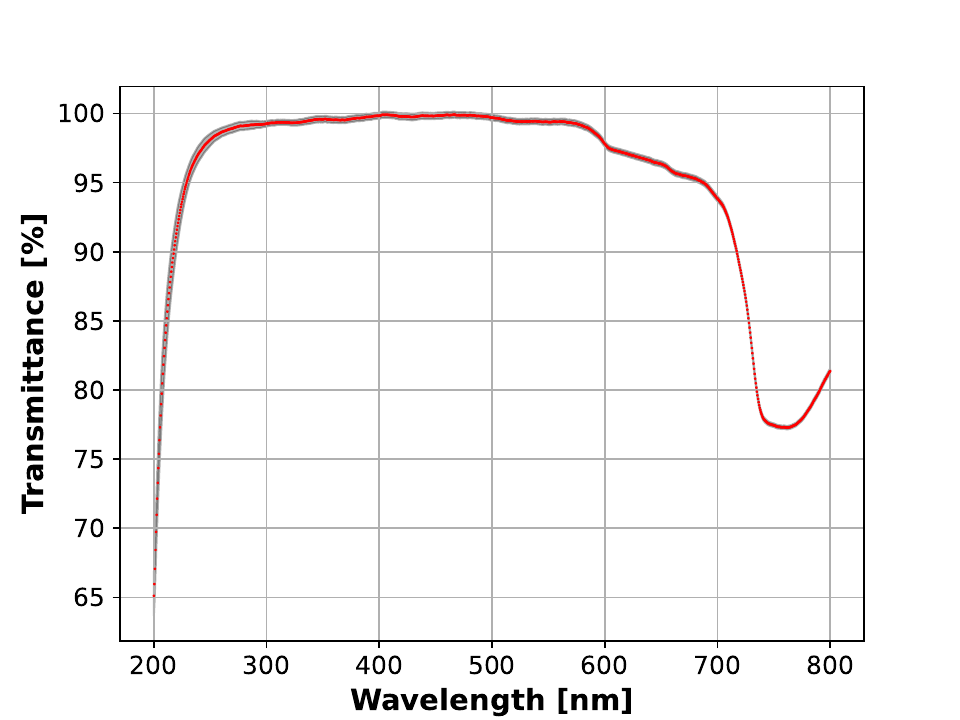}
    \caption{\label{fig:transmissionDI}\%T of PocketWATCH water when running in pure water mode. Estimated systematic errors on the spectrum measurements are also shown.}
    \end{minipage}
    \hfill
    \begin{minipage}[t]{0.49\textwidth}
    \centering
    \includegraphics[width=\textwidth]{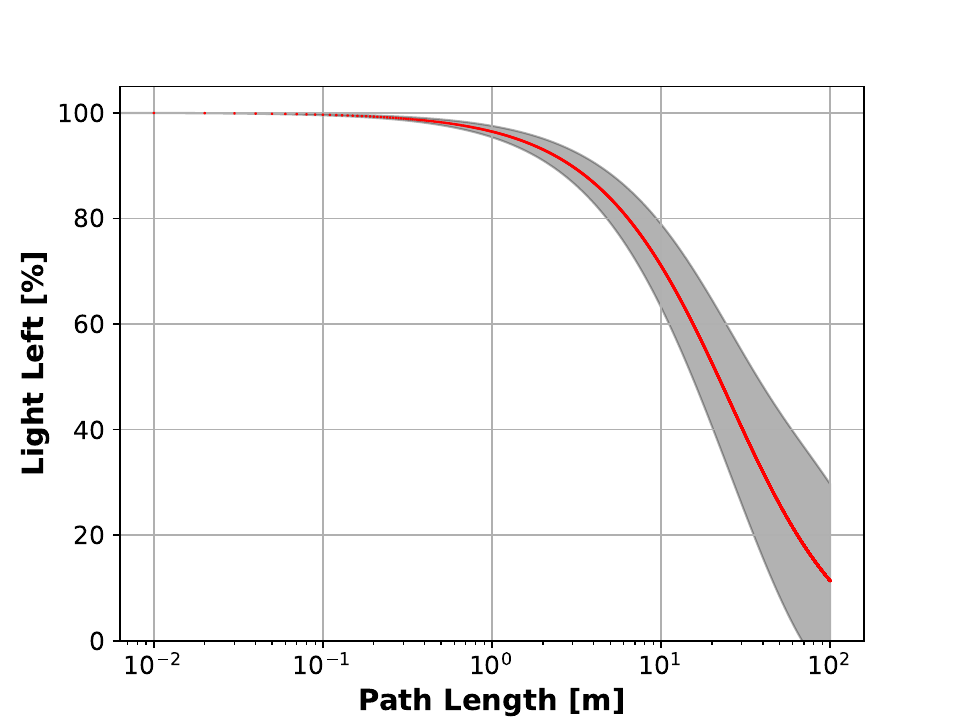}
    \caption{\label{fig:lightleftDI}\%LL($X$~m) shown as a function of path length in PocketWATCH pure water.}
    \end{minipage}
\end{figure}

The most relevant water quality metric for a very large water Cherenkov particle detector is its ability to transmit Cherenkov light to the photosensors, which are sensitive to mainly UV and visible light. To measure the UV transmission, we use a Shimadzu UV-2700i UV-Vis spectrophotometer with a 9~cm test cell. This test cell length is achieved by using a sample-filled 10~cm cell in the sample beam and a sample-filled 1~cm cell in the reference arm of the spectrophotometer. This measurement setup removes the overall effect of UV quartz cuvette transmittance and differing electromagnetic properties of the air-quartz and water-quartz boundaries, but is susceptible to differences in transmittance between the sample and reference quartz cells or cell cleanliness. An example of a transmission spectrum is shown in Fig.~\ref{fig:transmissionDI}. 

A useful metric for summarising the UV light transmission is in use by the SK experiment. The light left at $X$~m, \%LL($X$~m) as described in~\cite{MARTI2020163549}, is determined by first normalising the product of the Cherenkov emission spectrum and the PMT quantum efficiency spectrum to determine a spectrum of weights. For the Hamamatsu R3600-05 20\textquotedbl\hfill PMTs in SK, these weights are calculated and displayed in Table~\ref{tab:weights} for a selection of test wavelengths. Then a weighted average of the water \% transmittance (\%T) at various test wavelengths is calculated and scaled to a path length of $X$~m from the test cell length of 9~cm. 

\begin{table}[htb]
    \centering
    \begin{tabular}{@{}lccccccc@{}}
    \toprule
    Wavelength (nm) & 337 & 375 & 405 & 445 & 473 & 532 & 595 \\ \midrule
    Weight & 0.252 & 0.253 & 0.206 & 0.141 & 0.106 & 0.039 & 0.003 \\ \bottomrule
    \end{tabular}
    \caption{Weights calculated at various test wavelengths for the calculation of \%LL($X$~m).}
    \label{tab:weights}
\end{table}

For PocketWATCH, the light left at different path lengths is shown in Fig.~\ref{fig:lightleftDI}. Over a distance of 2~m, the weighted light left is observed to be $93\pm2$\%, which is sufficient to detect Cherenkov light from across the largest dimension of PocketWATCH using a PMT with similar quantum efficiency as in SK. For a path length of 15~m, the \%LL(15~m) is $62\pm10$\%. For comparison, the SK ``ultra-pure water'' range is from $\%\text{LL}(15\text{m})=75-82$\%, which was regularly achieved during the SK-III and SK-IV runs~\cite{Sekiya_2016}. 

There are several limitations in the method for measuring the transmittance of PocketWATCH ultra-pure water using the UV-Vis spectrophotometer. First, the instrument reaches maximum sensitivity on the range 350-550~nm. Because the test cell is only 9~cm, light attenuation in this range of wavelengths is indistinguishable by the spectrophotometer if the water is pure enough, limiting the ultimate sensitivity of the measurements. Second, there are several systematic uncertainties including the effects of cuvette cleanliness, cell positioning repeatability, and the handling of water samples between the plant room and the UV-Vis spectroscopy laboratory. The cumulative effect of systematic uncertainty of the method is quantified by taking the standard deviation of repeated measurements of \%T spectra of PocketWATCH water running in normal, ultra-pure mode over the course of several months. 

The effects of cuvette cleanliness were handled by following a rigorous procedure of rinsing with deionised water and cleaning with acetone and nitric acid, and by applying corrections to the data based on transmission measurements of the cuvettes when empty. Furthermore, the two sets of cuvettes are always paired the same way so that inherent optical differences in the quartz is consistent between measurements and effectively corrected.

To mitigate the effect of the systematic errors, a future improvement in the UV transmittance measurement procedure is planned. This will involve trading the fixed cells for a flow-through setup to allow continuous measurements while preventing the samples under measurement from being in contact with the atmosphere.

\section{Gadolinium-loaded Water Compatibility}
\label{sec:gdwater}

Since PocketWATCH was designed to be a test bed for future water Cherenkov neutrino and nucleon decay detection experiments which may employ gadolinium-loaded water, PocketWATCH is required to be compatible with a solution of 0.2\% Gd$_2$(SO$_4$)$_3\cdot$8H$_2$O and ultra-pure water. Using the experience of the EGADS experiment, construction materials were chosen to be specifically compatible and resistant to degradation. Most materials that are known to be compatible with ultra-pure water are known to be compatible with the desired gadolinium concentration, with notable exceptions of nylon, various rubbers and mastics, and certain proprietary PVC formulations~\cite{MARTI2020163549}.

Furthermore, there are two proven methods of selectively removing all dissolved ions except Gd$^{3+}$: bandpass filtration~\cite{MARTI2020163549}, and cation exchange resins with Gd counterions~\cite{ABE2022166248}. For PocketWATCH, however, since the requirements on \%T are not as strict as SK, EGADS or HK due to the smaller maximum path length of light in the tank, we instead forego deionisation altogether when filtering gadolinium-loaded water. In this mode, only particulates larger than $0.05$~$\mu$m are filtered following UV sanitisation.

\begin{figure}[htb]
    \centering
    \includegraphics[width=0.8\textwidth]{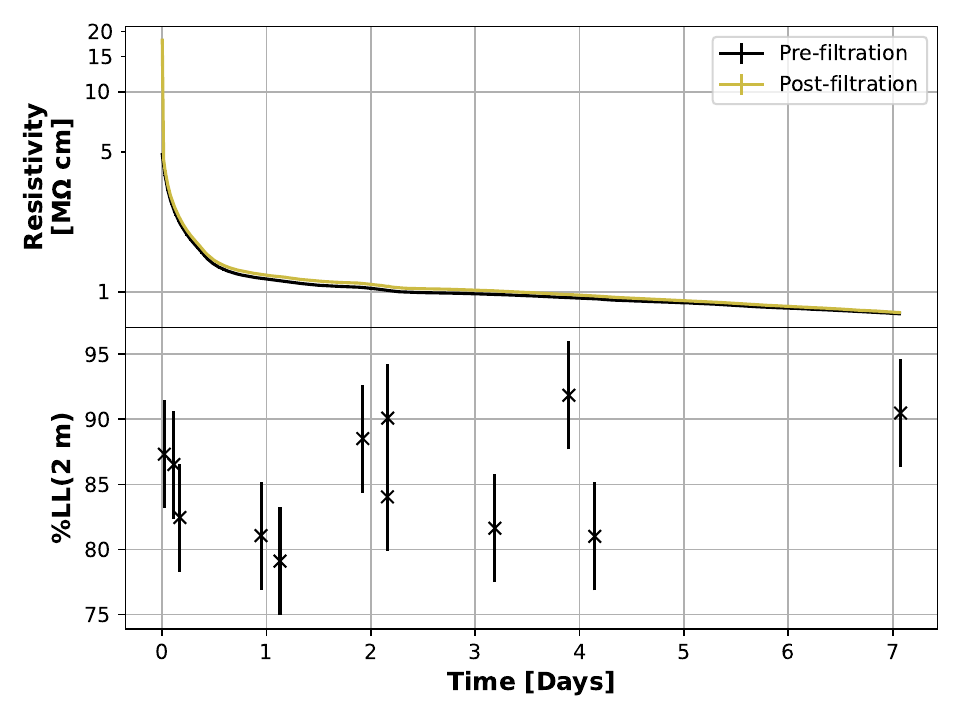}
    \caption{\label{fig:NoDI-old} Pure water resistivity (top) and \%LL(2 m) results (bottom) over time since the DI bypass. The mean value was \%LL(2 m) here is 85$\pm$5\%, but the standard cuvette cleanliness procedure was not followed. Instead, the stability of the \%LL(2 m) is of interest.}
\end{figure}

In this configuration, ions will gradually enter the water from various sources including atmospheric CO$_2$ and leaching of substances from wetted materials. The resistivity would thus decrease over a long period of time and would eventually reach some equilibrium. To observe these effects, we perform a test where the pure-water is exposed to the atmosphere and the DI cylinder is bypassed. Fig.~\ref{fig:NoDI-old} shows both the water resistivity and the measurements of \%LL(2 m) during this period. Even before Gd is loaded, the resistivity shows a rapid drop in the first day, and then decreases at a constant rate of approximately 9\% per day through the end of the test. During this time, no significant change in UV transmittance is measured so we are confident that the proposed filtration method is sufficient to maintain UV transmittance across the tank volume as required. 

We note that the \%T measurements which are used to calculate \%LL in Fig.~\ref{fig:NoDI-old} were taken prior to introduction of the nitric acid wash procedure for cleaning the UV cuvettes. These results, when compared with later measurements with the same water conditions (\%LL(2 m) = $93\pm2$\%), indicate that the cuvette cleanliness significantly affected the overall transmittance in the spectrophotometer measurements. However, the overall stability of the tank water \%T over the period of the DI bypass test is still demonstrated.

To test the filtration system in maintaining a useful environment for Gd-loaded water Cherenkov detector equipment testing, we dissolved 4~kg of Gd$_2$(SO$_4$)$_3\cdot8$H$_2$O into PocketWATCH to achieve a concentration of 0.2\% w/w. This is the desired concentration to achieve 90\% neutron captures on Gd~\cite{PhysRevLett.93.171101}. Immediately following dissolution, we monitored the resistivity and UV transmittance over the course of several days.

\begin{figure}[htb]
    \centering
    \includegraphics[width=0.8\textwidth]{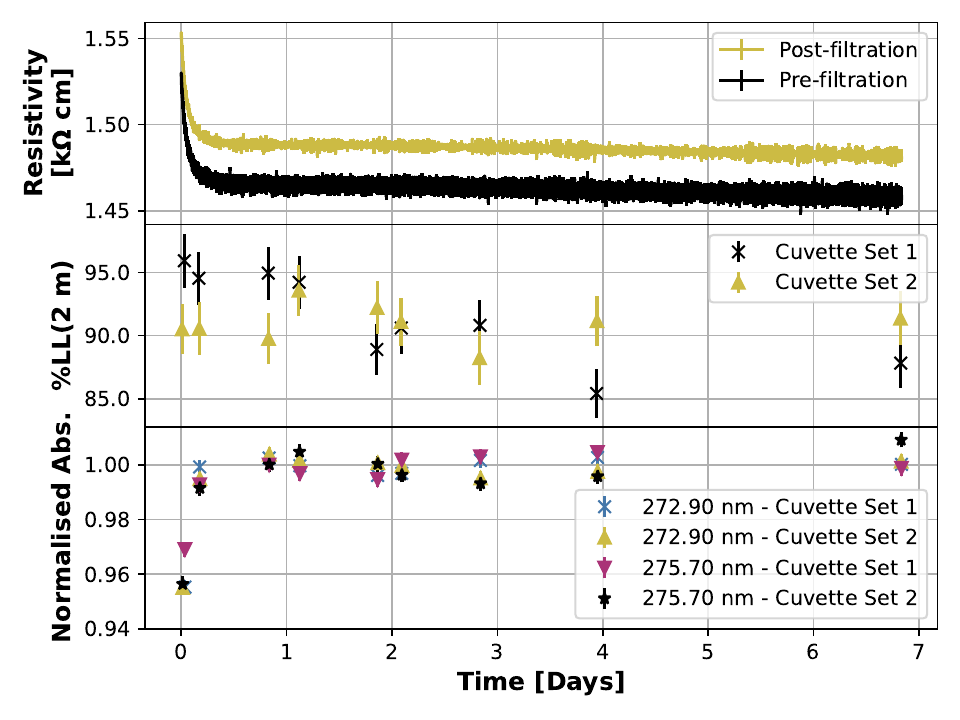}
    \caption{\label{fig:Gd} Gd water resistivity (top), \%LL(2 m) results (middle) and normalised peak absorbance for the largest two Gd absorption peaks (bottom). The approximately 1.5\% difference in resistivity readings from both UniCond sensors is consistent with the expected accuracy uncertainty of the cell constant, specified as $\pm 1\%$ for this range.}
\end{figure}

\begin{figure}[htb]
    \centering
    \includegraphics[width=0.7\textwidth]{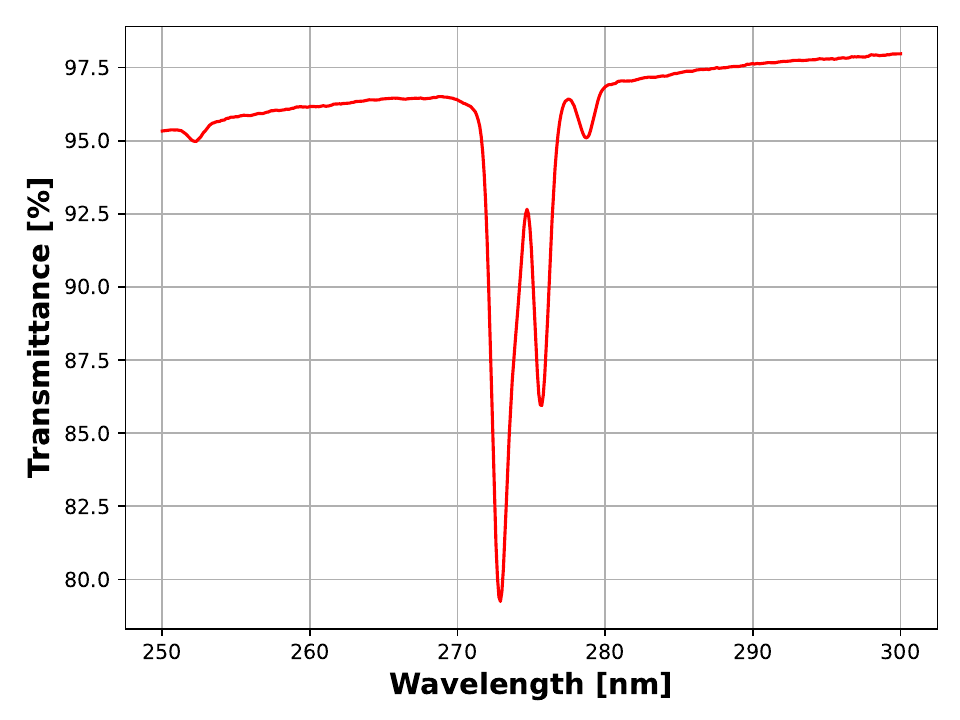}
    \caption{\label{fig:gdpeaks}Characteristic Gd absorption peaks.}
\end{figure}

The water resistivity dropped to approximately 1.5~k$\Omega\cdot$cm once the Gd was fully dissolved, then gradually fell by approximately 0.1\% per day due to dissolution of atmospheric gases and leaching (Fig.~\ref{fig:Gd} top). The measure of UV transmittance, similar to the DI bypass test, did not show a marked decrease from the pure water transmittance. The mean \%LL(2m) in the first few days of Gd operation is $91\pm2$\% and did not decrease below 85\% at any point during the test (Fig.~\ref{fig:Gd} middle). To monitor the stability of Gd concentration, we observe the amplitudes of the characteristic Gd absorption peaks between 270-280~nm (e.g. Fig.~\ref{fig:gdpeaks}). The amplitudes of the two largest peaks at 272.9 and 275.7~nm are monitored over the Gd test, showing no observable change in Gd concentration (Fig.~\ref{fig:Gd} bottom).

\section{Discussion}
\label{sec:discussion}

PocketWATCH was designed to be an accurate small-scale replica of the environment of the Kamioka-based family of water Cherenkov neutrino and nucleon decay detectors: SK, EGADS, and HK. For these detectors, it is important to maintain an ultra-pure water environment which can transmit UV light very long distances. Also, since the gain of photomultipliers is dependent on operating temperature, precise control of water temperature is critical. Finally, complete light-tightness of all feedthroughs and perforations is needed to operate PMTs in single photon counting mode. PocketWATCH has achieved these requirements.

The most significant limitation to our understanding of the water purity is the systematic uncertainty of the UV-Vis spectrophotometric measurements. The cuvette cleanliness introduces significant uncertainty and variation between measurements. Moreover, as the water purity is close to maximal, the 9~cm test cell is generally too small to cause UV attenuation on a scale that is detectable by the spectrophotometer. Extrapolating the attenuation of the 9~cm test cell to 2~m or 15~m multiplies the uncertainty further. Finally, the procedure for making measurements will be soon improved by incorporating flow-through cells for automating transmittance measurements.

We can be confident, however, that the overall scale of UV transmittance is accurate. PocketWATCH, when operating with pure or Gd-loaded water, effectively transmits UV light along path lengths that are on the same scale as the tank dimensions. This makes the facility a useful test bed for using Cherenkov light as a light source for evaluating calibration hardware or PMT characteristics. Having operated since August 2018, PocketWATCH has demonstrated excellent stability and performance.

\section{Conclusion}
\label{sec:conclusion}

We have shown that PocketWATCH is suitable for performing tests of hardware in an accurately replicated environment as SK, EGADS, and what is anticipated for HK. The temperature control, water purity, and low-light environment make it an ideal \textit{ex situ} test bed for performing calibration hardware characterisation and PMT dark noise studies. The latter results were recently published~\cite{Wilson2023}, and HK calibration light source pre-calibration work is ongoing. The facility might be used for any \textit{ex situ} hardware development studies relevant to water Cherenkov particle detectors.

\section{Acknowledgements}
This work was supported by the U.K. Science and Technology Facilities Council [grant numbers ST/R000069/1, ST/T00200X/1, ST/V002821/1, ST/V006185/1, ST/S006400/1, ST/X002438/1].


\bibliographystyle{JHEP}
\bibliography{biblio.bib}

\end{document}